\documentclass[
 reprint,
 nofootinbib,
 showkeys,
 amsmath,
 amssymb,
 aps,
 prb,
 floatfix,
 superscriptaddress,
 longbibliography
 ]{revtex4-2}

\usepackage{graphicx}
\usepackage{multirow}
\usepackage[dvipsnames]{xcolor}
\usepackage{dcolumn}
\usepackage{bm}
\usepackage{dsfont}
\usepackage{physics}
\usepackage[hidelinks, colorlinks=false, pdfborder={0 0 0}]{hyperref}
\usepackage{changepage}

\usepackage{ulem}

\usepackage[version=3]{mhchem} 
\usepackage[dvipsnames]{xcolor}
\usepackage{textcomp}
\usepackage{graphicx}
\usepackage{amsmath}
\usepackage{fixmath}
\usepackage{amsfonts}
\usepackage{amssymb}
\usepackage{upgreek}
\usepackage{bm}
\usepackage{ulem}

\begin{document}

\title{Optical probing of Wigner crystallization in monolayer WSe$_2$ via diffraction of longitudinal excitons}

\author{Artem N. Abramov}
\affiliation{School of Physics and Engineering, ITMO University, Saint Petersburg 197101, Russia}
\author{Emil Chiglintsev}
\affiliation{Moscow Institute of Physics and Technology (National Research University), Dolgoprudnyi, Moscow Region 141701, Russia}
\affiliation{Russian Quantum Center, Skolkovo, Moscow 143025, Russia}
\author{Tatiana Oskolkova}
\affiliation{School of Physics and Engineering, ITMO University, Saint Petersburg 197101, Russia}
\author{Maria Titova}
\affiliation{Laboratory of Programmable Functional Materials, Center for Neurophysics and Neuromorphic Technologies, Moscow 121205, Russia}
\affiliation{Moscow Center for Advanced Studies, Moscow, Russia}
\author{Mikhail Kashchenko}
\affiliation{Laboratory of Programmable Functional Materials, Center for Neurophysics and Neuromorphic Technologies, Moscow 121205, Russia}
\affiliation{Moscow Center for Advanced Studies, Moscow, Russia}
\author{Alexander Chernov}
\email{Corresponding author: a.chernov@rqc.ru}
\affiliation{Moscow Institute of Physics and Technology (National Research University), Dolgoprudnyi, Moscow Region 141701, Russia}
\affiliation{Russian Quantum Center, Skolkovo, Moscow 143025, Russia}
\author{Vasily Kravtsov}
\affiliation{School of Physics and Engineering, ITMO University, Saint Petersburg 197101, Russia}
\author{Ivan V. Iorsh}
\email{Corresponding author: ivan.iorsh@queensu.ca}
\affiliation{Department of Physics, Engineering Physics and Astronomy,
Queen’s University, Kingston, Ontario K7L 3N6, Canada}

\begin{abstract}
Monolayer transition metal dichalcogenides (TMDs) are characterized by relatively large carrier effective masses and suppressed screening of the Coulomb interaction, which substantially enhances the correlation effects in these structures. 
The direct band gap allows to effectively optically probe these correlations. 
Here, we present an experimental observation of Wigner crystallization in monolayer $\mathrm{WSe}_2$ probed by the measurement of the exciton diffraction on the Wigner crystal (WC) periodic potential. 
We observe the formation of the WC phase in the absence of external magnetic fields at temperature range $T<26~\mathrm{K}$ and carrier concentrations $n$ $<2\times10^{11}~\mathrm{cm}^{-2}$. 
The direct observation of the exciton diffraction is enabled by the strong exciton longitudinal-transverse splitting induced by the long-range intervalley exchange interaction, leading to the large detuning between main exciton peak and first diffraction peak. 
Our findings highlight that the valley degree of freedom of charge carriers in TMDs facilitates optical probing of correlated electron phases in these structures.
\end{abstract}

\maketitle
\noindent
\noindent
Mono- and few-layer Van der Waals (VdW) heterostructures proved to be an ideal platform to explore many-body correlations in electronic systems. 
Enhanced correlation effects are dictated by the suppressed Coulomb screening and relatively large effective masses. 
In recent years a plethora of strongly correlated electron phases have been predicted and observed in VdW heterostructures~\cite{cao2018correlated, chen2019evidence, cao2018unconventional, wu2019topological, xu2020correlated, regan2020mott, li2021continuous, lian2024valley, gao2024excitonic, cutshall2025imaging}, suggesting that these systems can be employed to perform quantum simulations~\cite{kennes2021moire}.

A paradigmatic example of the correlated electron phase is Wigner crystal~\cite{wigner1934interaction}.
Wigner crystallization occurs when the Coulomb repulsion between charge carriers begins to prevail over their kinetic energy.
This phase transition is governed by the dimensionless parameter r$_s=(a_B\sqrt{\pi n})^{-1}$, which is the ratio of the average distance between the charge carriers defined by concentration $n$ and the effective Bohr radius $a_B$. 
The system crystallizes when $\mathrm{r}_s$ exceeds the critical value $\mathrm{r}_s^*= 1$ \cite{drummond2009phase,tan2024importance}. 
The WC was first observed in a 2D electron gas on the surface of liquid helium~\cite{crandall1971crystallization, monarkha1975theory, grimes1979evidence}. 
It was further realized in  A$_3$B$_5$ quantum wells~\cite{andrei1988observation}  at temperatures $<100$ mK and magnetic fields exceeding $10$ T. 
Large magnetic fields are usually required to observe Wigner crystal, since electrons are localized at the scales of the magnetic length increasing the effective ratio of potential and kinetic energies.

Recently, WC was reported in both monolayer  \cite{smolenski2021signatures,sung2025electronic} and  bilayer TMDs~\cite{tsui2024direct,zhou2021bilayer} as well as in twisted moir\'e heterostructures~\cite{regan2020mott,li2024mapping,shimazaki2020strongly}. 
In the latter case, the observed phase is a generalized Wigner crystal with the period determined not only by the electron-electron interactions but by the periodicity of the moir\'e lattice.

There are several ways to probe the emergence of the WC phase. 
The Wigner crystal may become pinned by disorder or moir\'e potential (in the case of generalized moir\'e crystal) which significantly enhances the conductance of the sample compared to the Fermi liquid phase. 
Alternatively, one may probe the shear modes present in the crystalline phase with spatially modulated ac electric field in the MHz range~\cite{andrei1988observation}. 
Furthermore, the WC can be directly visualized via scanning tunneling microscopy~\cite{tsui2024direct}. 
TMDs offer an alternative optical method to probe WC: exciton-electron coupling induces a periodic potential for the exciton in WC phase, and the resulting exciton umklapp scattering leads to the redistribution of the exciton oscillator strength to the first diffraction peaks detuned by the energy associated with the WC reciprocal lattice vector. 
This method was used in~\cite{smolenski2021signatures} to measure the onset of WC phase in monolayer MoSe$_2$ characterized by the largest carrier effective mass among the TMD family.
However, the complex exciton landscape in 2D semiconductors makes it possible to use various approaches to detect emerging strongly correlated states.

In this Letter, we study Wigner crystallization in monolayer WSe$_2$ without external magnetic field and find previously unobserved features in the associated optical response.
Importantly, we expect to see the emergence of two branches of umklapp scattering. 
The strong long range intervalley exchange coupling of excitons in TMD enables strong longitudinal-transverse splitting, and the exciton dispersion is split into two branches: excitons with dipole moments perpendicular to the wavevector have the conventional parabolic dispersion, and the excitons with dipole moment parallel to the wavevector have much steeper linear dispersion. 
The umklapp scattering of the exciton branch with linear dispersion leads to the modified dependence of the diffraction peak on carrier concentration and to larger detunings from the main exciton peak, which allows us to accurately extract the WC associated peak position and corresponding oscillator strength. 

\begin{figure}[!h]
	\includegraphics[width=0.93\columnwidth]{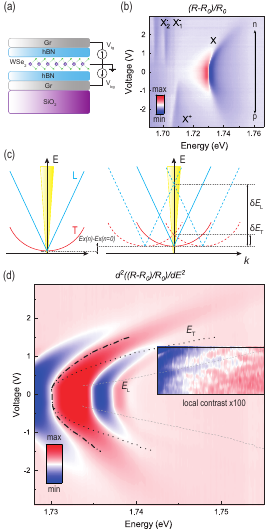}
	\caption{\textbf{Optical detection of Wigner crystallization.} 
	(a) Schematic of the studied device, consisting of WSe$_2$ monolayer, layers of hBN, graphene contact and gates with connected voltage sources. (b) Dependence of the reflectance contrast spectrum on the gate voltage, with X$^0$ corresponding to main exciton, X$^+$ and X$^-$ corresponding to positive and negative trions, the arrow indicates electron (n) and hole (p) doping. (c) The left panel shows the excitonic branches in monolayer WSe$_2$ with parabolic (red) and linear (blue) dispersion. The right panel shows the change of main exciton energy due to doping and the folding of the energy bands due to the Wigner crystal potential and the resulting diffraction peaks for both exciton branches. (d) Second derivative of reflection spectra with respect to energy. The contrast is increased by 100 times in the selected area. The map shows the blueshift of main exciton extracted from fit (dash-dotted line) and energy shift of Wigner resonances (the excitonic branch with parabolic dispersion is denoted by black dotted line, branch with linear dispersion is denoted by grey dashed line).}
	\label{fig:Setup}
\end{figure}

The geometry of the fabricated structure is shown in Fig.~\ref{fig:Setup}(a). 
The WSe$_2$ monolayer was encapsulated in atomically smooth hexagonal boron nitride (hBN).
Two gates (top and bottom) were implemented in the structure for precise control of charge carrier density $n$ and the displacement electric field. 
The electrical contacts and gates were made of graphene.
The carrier density was tuned by applying voltage to a single gate or to both gates simultaneously (Voltage = V$_{tg}$ + V$_{bg}$).
Throughout the measurements, the polarities of $V_{tg}$ and $V_{bg}$ were always kept the same, while the magnitude and sign of Voltage were varied from negative to positive values to achieve consistent electron or hole doping.

Fig. 1b shows the typical dependence of the reflection spectrum of the WSe$_2$ monolayer on the applied voltage at a temperature of 8 K. 
Herewith, we determine the value of the neutrality point at a voltage corresponding to the minimum energy of the main exciton. 
The dependence of the reflection spectrum on voltage demonstrates the blueshift of the exciton and the appearance of positive trion X$^{+}$ and negative trion X$^{-}$ as voltage departs from the neutrality point. 
We use the reflection data to directly map the applied voltage to the charge carrier density~\cite{glazov2020optical,smolenski2019interaction} (see SI). 

While no additional features are visible in the reflection map in Fig.~\ref{fig:Setup}(b), it is expected that formation of the Wigner crystal should lead to additional higher-energy Fano-like peaks. 
Their origin is schematically depicted in Fig.~\ref{fig:Setup}(c): in the translationally invariant system, there are two exciton branches split by the intervaley long-range exchange interactions. 
For wavevectors outside the light cone, these correspond to the excitonic branch with parabolic dispersion which has the dipole moment perpendicular to the wavevector (marked with "T") and the branch with almost linear dispersion "L" with the dipole moment parallel to the wavevector. 
In the WC phase, the exciton experiences the crystal periodic potential with the period defined by the carrier concentration. 
The umklapp scattering induces diffraction of excitons with momenta equal to the reciprocal lattice vectors to the $\Gamma$ point, where they can couple to photons. 
The oscillator strength is redistributed from the main exciton to this diffraction peaks, and their relative oscillator strength is proportional to the depth of the WC induced periodic potential. 
\begin{figure*}[!t]
	\includegraphics[width=0.85\textwidth]{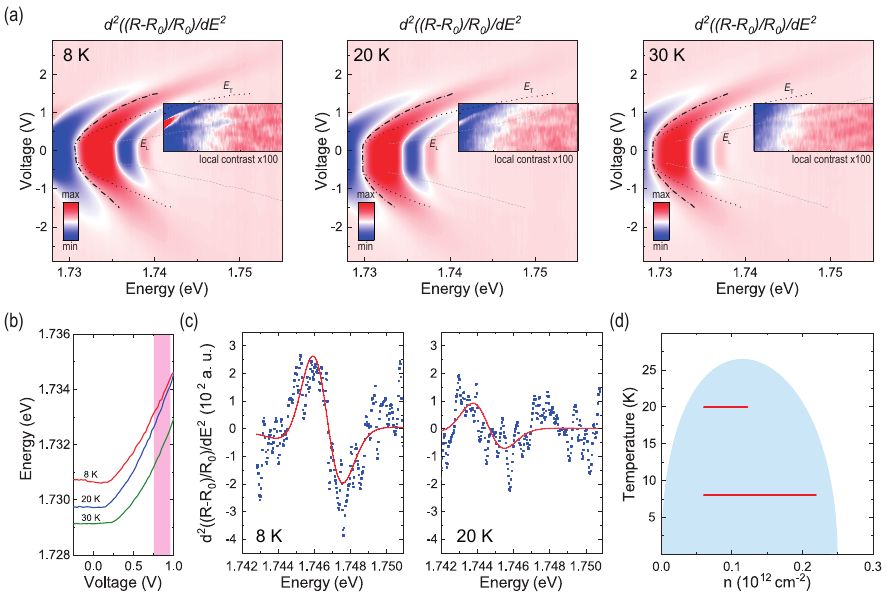}
	\caption{\textbf{Observation of Wigner crystal at different temperatures.} 
	(a) Dependence of the  reflectance contrast spectrum on the gate voltage for different temperatures (8~K - left, 20~K - center, 30~K - right). (b) Dependence of main exciton energy on applied voltage at a temperature of 8~K (red), 20~K (blue), 30~K (green). Pink area indicates the displacement of the main exciton corresponding to the range of carrier densities where Wigner crystal resonance is discernible. (c) Second derivatives of reflection spectra at 8~K and 20~K (blue squares) and fit functions (red solid lines). (d) Phase diagram. The blue region corresponds to the theoretically calculated state of the Wigner crystal. The red lines show the range of carrier densities at temperatures of 8~K and 20~K, where the Wigner crystal is observed experimentally.}
	\label{fig:Temp_dependance}
\end{figure*}

In Fig.~\ref{fig:Setup}(d) the map of the second derivative of the reflection is shown, and the weak feature in the energy range above 1.743 eV, which we associate with the first diffraction peak of the longitudinal exciton branch, is visible. 
Dash-dotted dot line corresponds to voltage-dependent spectral shift of the main exciton. 
Dashed and dotted lines in the plot correspond to the theoretical predictions for the energies of the first diffraction peaks for longitudinal and transverse excitons. 
They were calculated as follows. 

The energy offset between the main exciton peak and the first diffraction peak $\delta E_{T(L)}(V)$ at fixed gate voltage $V$ is given by
\begin{align}
\delta E_{T(L)}(V)= E_{T(L)}(V,G)-E_{T(L)}(V,0),
\end{align}
whe $E_{T,L}$ are the exciton energies of the two branches, and  $G$ is the absolute value of the reciprocal lattice vector, which for the case of the triangular lattice is expressed as $G=\sqrt{2\pi^2\sqrt{3}n}$, where $n=n(V)$ is the free carrier concentration. 
In the regime of not very strong doping, when the Fermi energy $E_F$ is much smaller than the trion binding energy $E_{tr}$, i.e. the energy difference of exciton and trion energies at zero carrier concentration, the free carrier induced exciton blueshift grows linearly with concentration~\cite{glazov2020optical}: 
\begin{align}
\delta E_X(n)=E_X(V,K=0)-E_X(V,K=0)|_{n=0}=\frac{n 2\pi\hbar^2}{m}
\end{align}
where $m=m_cm_X/(m_{tr})$ is the reduced mass expressed via the corresponding carrier effective mass, $m_c$, exciton mass $m_X=m_c+m_{\bar{c}}$, and trion mass $m_{tr}=2m_c+m_{\bar{c}}$, with $c=e,h$ for electron and hole doping, respectively.

Therefore, we can express the concentration, and thus, the energies of the diffraction peaks, via the main exciton blueshift.
The dispersion of the transverse and longitudinal excitons for the wavevectors lying beyond the light cone, where the retardation effects can be neglected, is given by~\cite{glazov2015spin}:
\begin{align}
&E_{T}(V,K)=E_X(V,K=0)+ \frac{\hbar^2}{(2m_X)} K^2, \\&E_{L}(V,K)=E_X(V,K=0)+\frac{\Gamma_0}{E_X}\frac{c}{n_{eff}}K,
\end{align}
where $\Gamma_0$ is the radiative broadening of the exciton, and $n_{eff}\approx 1.7 $ is the effective refractive index of the monolayer's substrate and superstrates. 
Therefore, the expected energies for the first diffraction peaks for both exciton branches can be expressed via the voltage dependent main exciton blueshift. 
These energies are plotted in Fig.~\ref{fig:Setup}(d) for $m_e=0.4~m_0,\quad m_h=0.5~m_0$,
which are typical values for WSe$_2$ monolayer~\cite{kormanyos2015k, rasmussen2015computational}. 

We see that the predicted energy of the diffraction peak for the excitonic branch with linear dispersion fits closely to the weak feature observed in the experimental map. 
At the same time, the predicted energy of the transverse exciton diffraction peak significantly overlaps with the tail of the main exciton peak, making it challenging to discern in the experimental spectra (see SI).

In Fig.~\ref{fig:Temp_dependance}(a) we show the temperature evolution of the first diffraction peak for the excitonic branch with linear dispersion. 
As can be seen, as temperature increases, the contrast of the feature decreases, and it eventually disappears in the 20-30~K range. 
Moreover, the feature is present only in the relatively narrow range of carrier concentrations equivalent to blueshift of the main exciton shown in Fig.~\ref{fig:Temp_dependance}(b). 
The red, blue and green lines trace the blueshift of the main exciton under applied voltage at 8, 20 and 30 K, respectively. 
The pink area marks the region where this feature is observed.
In order to extract the energies and relative oscillator strengths of the exciton diffraction peaks, we fitted the second derivative of the reflection contrast with respect to the energy in the vicinity of the feature with the second derivative of a Fano lineshape (see SI). 
Typical second derivatives of the reflection contrast for temperatures of 8 K and 20 K (blue squares) and the corresponding second derivative of a Fano lineshape (red lines) are shown in Fig.~\ref{fig:Temp_dependance}(c).
Using the amplitude of the observed diffraction peak at 8~K and 20~K, and noting its absence at 30 K, a linear extrapolation shows that the amplitude vanishes at a temperature of 26~K, indicating a phase transition at this temperature.

In Fig.~\ref{fig:Temp_dependance}(d) we plot the theoretically calculated phase diagram of monolayer WSe$_2$ and mark regions where the WC crystallization was observed experimentally with red lines. 
To evaluate the phase diagram, we use the seminal result of Platzman and Fukuyama~\cite{PhysRevB.10.3150} stating that the melting curve of the two dimensional electron liquid can be approximated by the equation $\langle E_{Coulomb} \rangle/\langle E_{kin}\rangle =\gamma$ where $\langle E_{Coulomb} \rangle$ and $\langle E_{kin}\rangle$ are the average Coulomb interaction and kinetic energies, and $\gamma$ is the material specific parameter which generally is $\gg 1$. 
For example, for electrons on the surface of liquid helium $\gamma \approx 130$. 
Here, we chose $\gamma$ such that the highest phase transition temperature corresponds to the largest temperature of 26 K where the WC phase was detected. 
This yields to $\gamma\approx 25$, which corresponds to $r_s^*\approx 20$. 
This value $r_s^*$ is lower than the value $30$ predicted for the WC in monolayer TMDs in~\cite{PhysRevB.95.115438}. 
Specifically, fit of experimental data corresponds to the maximal density for which WC phase exists  at zero temperature equal to $2.5\times 10^{11}$~cm$^{-2}$ which is  at least two times larger than the one predicted theoretically for TMD systems. 
We attribute this discrepancy to the presence of disorder. 
Indeed, as was recently shown in~\cite{joy2025disorderinducedliquidsolidphasecoexistence} (see SI) charged impurities  can substantially increase the critical density with respect to that in the clean system. 
We also anticipate that the effect of charged impurities might explain the fact that we only see the WC at electron doping but not for the hole one (this asymmetry was also observed for the case of MoSe$_2$ in~\cite{smolenski2021signatures}). 
Different samples have different relative concentration of the defect states in the vicinity of conduction and valence band, and therefore the impurities have different effect in the hole- and electron- doped regions. 

The Wigner crystal-induced periodic potential drives the redistribution of oscillator strength from the main exciton to the diffraction peak of the excitonic branch, making the strength of this diffraction peak a direct quantitative measure of the potential.
Under the approximation of the weak periodic potential, the relative oscillator strength of the first diffraction peak with respect to the main one yields~\cite{voronov2003resonance}:
\begin{align}
f_{T(L)}\approx   \frac{\sum_{i}|V_{G_i}|^2}{|\delta E_{T(L)}|^2},
\end{align}
where $V_G$ is the Fourier component  of the periodic potential induced by the WC evaluated at the six reciprocal lattice vectors, closest to the $\Gamma$ point. 
The electron density in the WC phase can be approximated by $n(\mathbf{r})=\sum_{m,n} (\sqrt{\pi} u)^{-2} \exp[-(\mathbf{r}-m\mathbf{a}_1-n\mathbf{a}_2)^2/u^2]$, where $\mathbf{a_1},\mathbf{a_2}$ are lattice vectors of the triangular lattice, and $u$ is the mean variance of the position of the electron in the unit cell. 
We assume that the induced potential can be approximated by $V(\mathbf{r})=\partial V/\partial n|_{n=0} n(\mathbf{r})$. 
The Fourier component of the potential then yields:
\begin{align}
|V_{G}| \approx \delta E_X(n) e^{-\pi^2 u^2/a^2},
\end{align}
where $a$ is the WC period. 
The mean variance of the electron position in the unit cell can be approximated by~\cite{monarkha2012two}:
\begin{align}
u^2/a^2\approx \frac{0.463}{\sqrt{r_s}}+0.75 r_s \frac{T}{e^2/a_B}\ln \left(\frac{L}{a_B}\right). \label{eq:Displacement}
\end{align}

\begin{figure}[!h]
	\includegraphics[width=1\columnwidth]{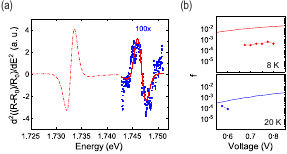}
	\caption{\textbf{Oscillator strength of the longitudinal exciton diffraction peak.} 
	(a) Second derivative of main exciton resonance fit function (red dash-dotted line) and experimental second derivative of the Wigner resonance spectra (blue squares) and its fit function (red solid line). Values are increased by 100 times for the diffraction peak. (b) Theoretically calculated dependence of the relative oscillator strength on charge carrier density (solid lines) and experimental values (circles) for 8~K (red data) and 20~K (blue data).}
	\label{fig:Oscillator}
\end{figure}

The first term in Eq.~\eqref{eq:Displacement} comes from the quantum fluctuations and is present at zero temperature. 
The second term comes from the thermal fluctuations and corresponds to the excitation of the long wavelength phonon mode. 
It diverges logarithmically in the limit of of the infinite sample, i.e. when the system lateral dimension  $L\rightarrow \infty$ as is usual for two dimensional systems. However, for the experimental system size of $L\approx 2~\mu$m the thermal contribution is finite and within the experimental parameter range $u/a < 0.3$.
We extracted the second derivative of the reflection contrast associated with the diffraction peaks (see SI) and fit it with the Fano lineshape.
Fig.~\ref{fig:Oscillator}(a) shows a typical second derivative curve of the reflection contrast in the region of the diffraction peak for the linearly dispersing excitonic branch (blue squares) at a gate voltage of 0.725 V and a temperature of 8 K, along with the corresponding Fano lineshape fit (red solid line). 
For comparison, the Figure also displays the second derivative of the reflection contrast for the main exciton under the same voltage and temperature conditions (red dash-dotted line), calculated using the Fano resonance equation with parameters extracted from fitting the reflection contrast of the main exciton.
We have obtained values of the oscillator strength ratio between the diffraction peak and main exciton peak for the linearly dispersing excitonic branch $f$ with $\sim$ $3\times10^{-4}$ for 8 K and $9\times10^{-5}$ for 20 K. 
We estimate the uncertainty of the extracted value by about an order of magnitude, which comes predominantly from uncertainty in extracting the radiative decay rate of the linearly dispersing excitonic branch from the second derivative of the reflection contrast. 
In Fig.~\ref{fig:Oscillator}(b) we plot the theoretical dependence of the relative oscillator strength on the gate voltage (curves) and compared it to the values extracted from the experiments (dots). 
As can be seen, while for the high temperatures (20~K, lower panel) the theoretical estimation agrees well with the experimentally observed value, the theoretical prediction for low temperature (8~K, top panel) overestimates the experimental relative oscillator strength by an order of magnitude. 
We attribute this discrepancy to the disorder which leads to the additional variation of the electron position in the unit cell and suppresses the experimentally observed contrast.

To conclude, in this Letter we report, for the first time, the optical observation of Wigner crystallization in monolayer WSe$_2$.
We optically observe a diffraction peak of the linearly dispersing excitonic branch, caused by the Wigner crystal potential in the absence of the external magnetic field. 
We highlight the importance of the intervalley exchange coupling in the monolayer for the optical observability of the Wigner crystallization, since the longitudinal exciton branch, which has a significantly steeper dispersion than the transverse one, allows us to spectrally resolve the main exciton and first diffraction peak resulting from the umklapp scattering. 
Our results suggest that presence of disorder in experimental samples is an important factor for the observability of the effect as was predicted in~\cite{joy2025disorderinducedliquidsolidphasecoexistence}, since the concentration and temperature range for which we observe the WC phase exceed those predicted theoretically for the clean monolayers~\cite{PhysRevB.95.115438}. 
We believe that the results presented highlight that the interplay of valley physics and exciton effects make the monolayer TMD a powerful platform for the optical probing of strong correlation effects.

This study was supported by Priority 2030 Federal Academic Leadership Program. 
T. O. and V. K. acknowledge support from Russian Science Foundation project 25-42-01019.
The work was supported by Rosatom in the framework of the Roadmap for Quantum computing (Contract № 868/2054-D  dated 12 December 2025).

\newpage

\section*{Supplementary Information}

\section{Measurement of the dependence of reflection spectra on the doping level}
We studied the fabricated sample using the following optical scheme. 
The sample was placed in a closed-cycle helium cryostat on a temperature-controlled platform with a base temperature of 7 K. 
For reflection measurements, the sample was illuminated with light from a picosecond pulsed supercontinuum laser operating in the 700-750 nm spectral range.
Illumination and collection of the reflected signal were carried out using a 50x objective with a numerical aperture of N.A. = 0.65. 
The objective was mounted on a three-axis piezo stage for precise positioning and focusing. 
The collected radiation was dispersed by a monochromator equipped with a 1200 g/mm grating and detected by a low-noise nitrogen-cooled CCD matrix.
Electrical contact to the sample was made to a two-channel source-meter unit for precise application of gate voltages and monitoring of leakage currents.
To analyze the reflection spectra, we first normalized each spectrum to the intensity at a reference wavelength far from the exciton resonance.
We then calculated the reflection contrast as (R–R$_0$)/R$_0$.
The reference spectrum R$_0$ was measured at the same location on the sample as the voltage-dependent series. 
To obtain R$_0$, a sufficiently high voltage was applied to the sample to cause a Fermi level shift that quenched the main exciton resonance, effectively suppressing the resonant contribution to the reflection.
Next, we fitted the reflection contrast spectra near the main exciton resonance using a Fano model for the resonant reflection amplitude. The model function for the complex reflection amplitude is given by:
\begin{align}
F(\omega) = \sqrt{R_{bg}} \exp(-i \phi_0) + \frac{i \Gamma_0}{(\omega - \omega_0) - i (\Gamma_0 + \Gamma_m)}
\end{align}
where R$_bg$ is the non-resonant background reflection intensity, $\phi_0$ is a phase shift, $\omega_0$ is the resonant frequency, $\Gamma_0$ is the radiative decay rate, and $\Gamma_m$ is the non-radiative decay rate. The modeled reflection contrast is then:
\begin{align}
\frac{R - R_0}{R_0} = \frac{|F|^2 - R_{bg}}{R_{bg}}
\end{align}
A typical reflection contrast spectrum of the main exciton at charge neutrality and its corresponding fit are shown in Supplementary Figure 1a. 
This fitting procedure allowed us to extract the voltage dependencies of the resonance frequency $\omega_0$(V) and the decay rates $\Gamma_0$(V) and $\Gamma_m$(V). 
Examples of these dependencies for one measurement series are shown in Figures S1b-d. 
The data demonstrate a monotonous evolution of these parameters with gate voltage, which suggests a relatively uniform sample quality in the probed region and consistent electrostatic tuning.
\begin{figure*}[!ht]
	\includegraphics[width=1\textwidth]{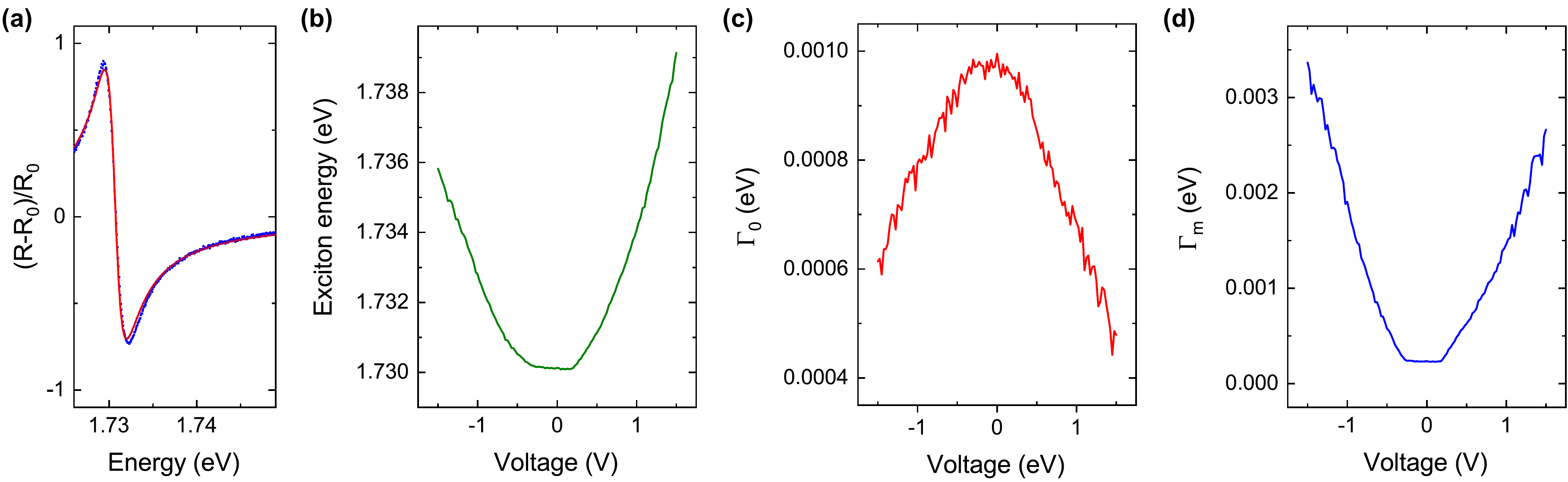}
	\caption{\textbf{Exciton resonance parameters.} 
	(a) The reflection spectrum of the main exciton at 8 K and at 0 V (blue squares) and its fit (red line). (b) Energy shift of the main exciton at applied voltage. (c) Change of $\Gamma_0$ at applied voltage. (d) Change of $\Gamma_m$ at applied voltage.}
	\label{fig:Temp_dependance}
\end{figure*}

\section{Determination of charge carrier concentration}
The concentration of charge carriers $n$ was estimated from the measured blueshift of the main exciton resonance, $\delta E_X$. 
To this end, we used the following relation, which is valid in the regime of low doping where the Fermi energy is much smaller than the trion binding energy
$\delta E_X=2(\pi\hbar^2/m)n$, where $m=m_cm_X/(m_{tr})$ is the reduced mass expressed via the corresponding carrier effective mass, $m_c$, exciton mass $m_X=m_c+m_{\bar{c}}$, and trion mass $m_{tr}=2m_c+m_{\bar{c}}$, where $c=e,h$ for electron and hole doping respectively. 
In this work, we used the effective mass values $m_e$ = 0.4 $m_0$ and $m_h$ = 0.5 $m_0$.
Supplementary Figure 2 shows a typical dependence of the electron concentration on the applied voltage to the sample at a temperature of 8~K.

\begin{figure}[!ht]
	\includegraphics[width=0.5\textwidth]{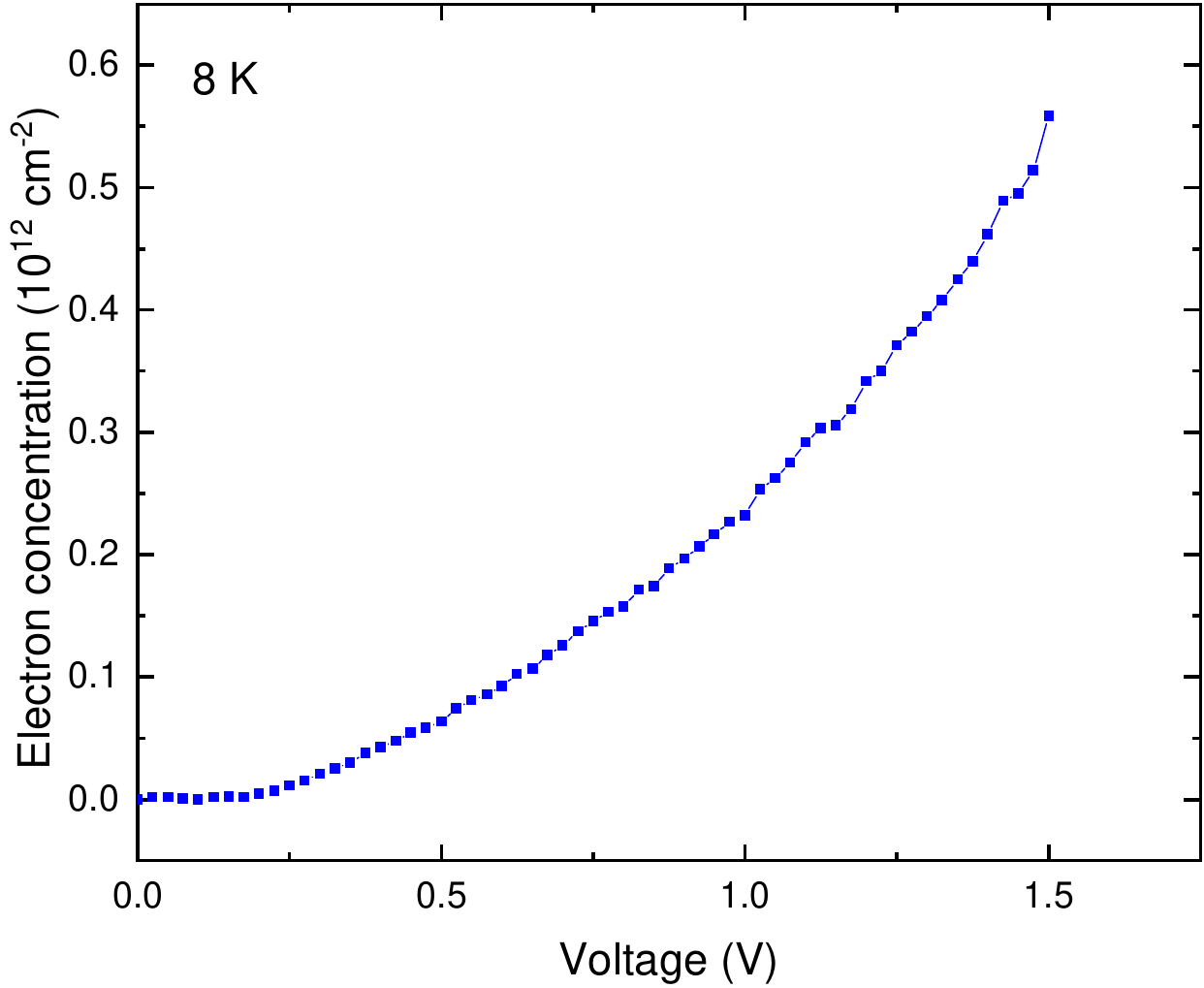}
	\caption{\textbf{Typical dependence of concentration on applied voltage.} 
	}
	\label{fig:Temp_dependance}
\end{figure}

\section{Differentiation of reflection spectra and observation of exciton diffraction}
The weak oscillator strength of the exciton diffraction feature prevents its reliable observation in standard reflection contrast maps of the WSe$_2$ monolayer.
 To enhance the visibility of these weak features, we analyzed the second derivative with respect to energy of the reflection spectra.
 A smoothing procedure was applied after each differentiation step to reduce noise. Furthermore, at each gate voltage, we subtracted the mean value of the second-derivative signal to compensate for minor artifacts arising from slight variations in the laser intensity, which were amplified by the differentiation process.
Given the very small amplitude of the observed resonances, we performed a statistical evaluation of our results. 
A feature related to Wigner crystallization was observed in 14 independent measurement series at a temperature of 8 K.
This effect was observed irrespective of whether the doping was applied using the top gate, the bottom gate, or both gates simultaneously.
Furthermore, we confirmed the thermal fading of the feature: the exciton diffraction signal attenuated and disappeared as the temperature was increased. It was absent in measurements at 30 K (5 measurements), 40 K (3 measurements), and 50 K (1 measurement).
We also note the probable observation of exciton diffraction associated with the transverse branch (parabolic dispersion) under hole doping conditions. 
The temperature evolution of this feature is presented in Supplementary Figure 3, where it is visible as a region of enhanced contrast in the second-derivative map.
The emergence of this resonance is consistent with theoretical expectations, as the larger effective mass of holes compared to electrons should facilitate Wigner crystallization.
However, the small energy separation between this diffraction peak and the main exciton resonance complicates its unambiguous identification and quantitative analysis, warranting further detailed investigation.

\begin{figure*}[!ht]
	\includegraphics[width=0.85\textwidth]{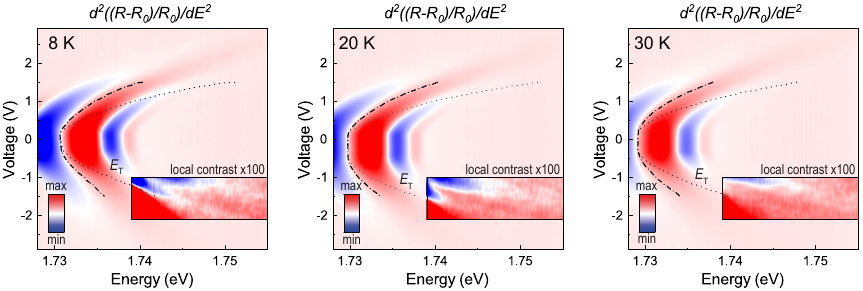}
	\caption{\textbf{Observation of Wigner diffraction for the exciton branch with parabolic dispersion.} 
	The Wigner resonance is observed in the marked area with 100 times increased local contrast. The Wigner resonance quenching is also observed with increasing temperature. Dash-dotted line indicates the displacement of the main exciton with increasing concentration of charge carriers. Dotted line represents the energy of the Wigner resonance.}
	\label{fig:Temp_dependance}
\end{figure*}

\section{Data Analysis for Extracting Wigner Resonance Parameters}
To quantify the weak characteristics associated with the resonance of the Wigner crystal in the second-derivative spectra, we have developed our own fitting procedure.
In our approach, we took into account that the signal from the WC diffraction peak is inherently weak and appears against a strong structured background resulting from the second derivative of the main exciton resonance. 
In addition, we expect that the line shape is affected by interference effects between the WC resonance, the main exciton resonance, and the nonresonant background, as well as potential inhomogeneous broadening.
The experimental second-derivative spectrum in the vicinity of the WC feature contains a dominant background slope from the tail of the main exciton. 
First, we removed this background by subtracting a linear baseline determined from a linear fit to the spectral regions immediately flanking the WC resonance (in the range of 10~meV). 

Next, we model the WC resonance accounting for inhomogeneous broadening and interference.
We assume that the resonant frequency of the WC state, $\omega_W$, is not a single value but is distributed around a mean $\omega_W^0$ due to local potential fluctuations. This distribution is described by a normalized Gaussian function:
\begin{align}
p(\omega') = \frac{1}{s\sqrt{\pi}} \exp[-\frac{(\omega' - \omega_W^0)^2}{s^2}],
\label{eq:gaussian_distribution_SI}
\end{align}
where $\omega'$ is a possible resonance frequency and $s$ is the standard deviation, which we set to 1 meV based on the typical scale of disorder broadening observed in similar high-quality encapsulated samples.
The optical response at a given frequency $\omega$ for a specific resonance $\omega'$ is given by a Lorentzian lineshape $\mathcal{L}(\omega')$, modified to include interference with a non-resonant background:
\begin{align}
\mathcal{L}(\omega') = \frac{i f \Gamma_0}{(\omega - \omega') - i ( f \Gamma_0 + \Gamma_m)},
\label{eq:lorentzian_model_SI}
\end{align}
where $f$ is the ratio of the oscillator strengths of the WC resonance and the resonance of the main exciton, $\Gamma_0$ and $\Gamma_m$ are the radiative and non-radiative broadening parameters of the main exciton (and the WC resonance). 
The resulting signal, assuming that the resonance of the main exciton is far from the Wigner resonance is:
\begin{align}
R{\text{$_W$}}(\omega')  = |\mathcal{L}(\omega')|^2 + 2\Re[\mathcal{L}(\omega') e^{-i\phi_0}].
\label{eq:reflection_intensity_SI}
\end{align}
The final reflection contrast $\Delta R / R_0$ is the expectation value of $R_{\text{W}}(\omega')$ over the distribution of $\omega'$:
\begin{align}
(\Delta R / R_0)(\omega) = \int_{-\infty}^{\infty} R_{\text{$_W$}}(\omega') p(\omega') d\omega'.
\label{eq:final_integral_SI}
\end{align}
This integral was computed numerically over the range $\omega' \in [\omega_W^0 - 5s, \omega_W^0 + 5s]$.
We compared our model with experimental data.
To do this, we calculated the second derivative of the simulated reflection contrast $d^2(\Delta R / R_0)/d\omega^2$.
The parameters $\Gamma_0$, $\Gamma_m$, and $\phi_0$ for the main exciton were fixed to the values obtained from the independent Fano fit of the undifferentiated reflection spectrum at the same gate voltage.
We have selected the values of the Wigner resonance energy and the oscillator strengths ratio.
The standard deviation $s$ was fixed at 1 meV.
Typical fits obtained with this procedure are presented in Figures 2c and 3a of the main text.
The model provides a good fit to the experimental lineshape.
However, we note that the weak signal of the diffraction peak and the complexity of the processing procedure introduce a strong error in the values of the fitting parameters. 
So, we estimate the error of the extracted value of $f$ in the order of magnitude of $f$.
\section{Reflection coefficient in the presence of the Wigner phase}
We assume that the excitons are doubly degenerate (we neglect the purely excitonic TE-TM splitting) and the exciton polarization two-dimensional vector satisfies the equation:
\begin{align}
&\partial_t \mathbf{P}_X(\mathbf{r},t) = \nonumber \\&-i\left(-\frac{\hbar^2}{2 m_X}\nabla^2 +V_W(\mathbf{r})-i\Gamma \right)\mathbf{P}_X(\mathbf{r},t) + \Gamma_0\mathbf{E}({\mathbf{r,t}}), 
\end{align}
where $V_{W}(\mathbf{r})$ is the periodic potential induced by Wigner crystallization, $\Gamma$ is the non-radiative damping rate,  and $\mathbf{E}$ is the total electric field:
The formal solution for $P_{X}$ is the given by:
\begin{align}
&\mathbf{P}_X(\mathbf{r},\omega)=\Gamma_0 \int d^2\mathbf{r'} \sum_{n,\mathbf{K}} \frac{|\Psi_{n,\mathbf{K}}(\mathbf{r})\rangle\langle \Psi_{n,\mathbf{K}}(\mathbf{r'})|}{E_n(\mathbf{K})-\omega-i\Gamma} \mathbf{E}(\mathbf{r'},\omega) = \nonumber \\&\int d^2{\mathbf{r'}} \hat{\alpha} (\mathbf{r,r'}) \mathbf{E}(\mathbf{r'})
\end{align}
where $|\Psi_{n,K}\rangle$ are the Bloch wave solutions for the excitons in the periodic Wigner potential and can be expanded as:
\begin{align}
\Psi_{n,\mathbf{K}}(\mathbf{r})= \sum_{\mathbf{Q}} C_{n,\mathbf{K,Q}} e^{i(\mathbf{Q+K})\mathbf{r}},
\end{align}
where sum is taken over the reciprocal lattice of the triangular lattice formed by the Wigner crystal. 

To find the reflection coefficient from the structure we expand the total electric field as:
\begin{align}
&\mathbf{E}(z,\mathbf{r},\omega) =\nonumber \\& \mathbf{E}_0(z,\omega) + \int d^2{\mathbf{r'}}\hat{G}(z,0, \mathbf{r-r'},\omega) \int d^{2}\mathbf{r''} \hat{\alpha}(\mathbf{r',r''})\mathbf{E}(z=0,\mathbf{r''})
\end{align}
where $\hat{G}$ is the photonic Green's function, and we have used that the monolayer in terms of optical response is a zero thickness layer situated at $z=0$. Taking the Fourier transform over $x,y$ coordinates we obtain:
\begin{align}
&\mathbf{E}(z,\mathbf{q},\omega) = \mathbf{E}_0(z,\omega)\delta_{\mathbf{q},0}  \nonumber\\&+\hat{G}(z,0, \mathbf{q},\omega) \int  d^2{\mathbf{r'}}d^{2}\mathbf{r''} \hat{\alpha}(\mathbf{r',r''})e^{i(\mathbf{qr'-q'r''})}\mathbf{E}(z=0,\mathbf{q'})
\end{align}
The periodicity of the Wigner potential induces the coupling between field components with in-plane wavevector projections which differ by a linear combination of reciprocal lattice vectors. If we consider the normal incidence then ater some algebra we obtain for the reflected field:
\begin{align}
&\mathbf{E}(z\rightarrow-\infty,\mathbf{q}=0,\omega) = \nonumber \\&\hat{G}(-\infty,\mathbf{q}=0,\omega)\left[\hat{\alpha}^{-1} - \hat{G}(z=0,\mathbf{Q'},\omega)\delta_{\mathbf{Q',Q''}}\right]^{-1}_{0,0} \mathbf{E}_0 \label{eq:ref_gen}
\end{align}
where summation is taken over the reciprocal lattice vectors $Q$, and the expression in square brackets is the matrix in the space of the reciprocal lattice vectors. The matrix $\hat{\alpha}_{\mathbf{Q',Q''}}$ yields:
\begin{align}
\hat{\alpha}_{\mathbf{Q',Q''}} = \sum_{n}\Gamma_0  \frac{C_{n,\mathbf{Q'}}C^*_{n,\mathbf{Q''}}}{E_n-\omega-i\Gamma}
\end{align}
The resonances in the reflection coefficients are defined by the zeros of the determinant of the matrix in the square bracket. 

For the case of  weak Wigner potential $V_W(\mathbf{r})$ and trigonal symmetry we can account only for the first order diffraction from the six reciprocal lattice vectors closest to the $\Gamma$ point: $\mathbf{Q}=\{\pm \mathbf{Q}_1, \pm \mathbf{Q}_2, \pm (\mathbf{Q}_1+\mathbf{Q}_2)\}$, where $\mathbf{Q}_1 = \frac{4\pi}{\sqrt{3}a}(0,1), \quad \mathbf{Q}_2 = \frac{4\pi}{\sqrt{3}a}(\sqrt{3}/2,-1/2) $, which all have the same absolute value $Q= 4\pi/(\sqrt{3}a)$. In the vicinity of $\omega \approx \omega_0 +\hbar^2Q^2/(2m_X)$ there are six almost degenerate eigenstates, but only one, the fully symmetric one couples to the far field at $\Gamma$ point. the fraction of the $\mathbf{Q}=0$ in the symmetric mode is given by:
\begin{align}
C_{s,0} =  \frac{\sqrt{6}V_Q}{\hbar^2Q^2/(2m_X)},
\end{align}
where $V_Q$ is the Fourier transform of the Wigner periodic potential:
\begin{align}
V_Q = \frac{1}{\Omega_0}\int d^2\mathbf{r} V_{W}(\mathbf{r}) e^{i\mathbf{Q_1 r}},
\end{align}
where $\Omega_0$ is the unit cell area.

We note that bare polarizability is a scalar in the polarization subspace, since longtidunial and transverse exciton are degenerate. Dressing by the vacuum electromagnetic field lifts this degeneracy, and for the case of the weak Wigner potential and for the case when $Q\gg \omega/c$, the expression in the parenthesis in Eq.~\eqref{eq:ref_gen} is given by:
\begin{align}
&\left[\hat{\alpha}^{-1} - \hat{G}(z=0,\mathbf{Q'},\omega)\delta_{\mathbf{Q',Q''}}\right]^{-1}_{0,0} =\nonumber\\& \frac{C_{s,0}^2\Gamma_0}{2}\left(\frac{1}{E_T(Q)-\omega-i\Gamma}+\frac{1}{E_L(Q)-\omega-i\Gamma}\right)\times \hat{I},
\end{align}
where $\hat{I}$ is the unity matrix, and $E_T,E_L$ are the energies of the transverse and longitudinal exciton energies are given by
\begin{align}
&E_T(Q) = E_X + \frac{\hbar^2 Q^2}{2m_X}, \\
&E_L(Q) = E_T(Q) + \frac{\Gamma_0}{E_X} \frac{c}{n_{eff}} Q,
\end{align}
where $E_X$ is the exciton energy at $Q=0$ and $n_{eff}$ is the effective refractive index of the structure.

We note that both longitudinal and transverse exciton have the same relative oscilattor strength, and the relative oscillator strength $f$ is $C_{s,0}^2/2$ yielding:
\begin{align}
f_{T(L)} = \frac{3V_Q^2}{|E_{T(L)}(Q)-E_X|^2}.
\end{align}

In order two find $V_Q$ we recall that the Wigner crystal induced potential comes from the interaction of the excitons with local variations of the electron density. Thus, the potential Fourier component $V_Q$ is given by:
\begin{align}
V_Q = \delta E_X(n) \int _{\Omega_0} d^2 \mathbf{r} n(\mathbf{r}) e^{i\mathbf{Q}_1\mathbf{r}},
\end{align}
where $n(\mathbf{r})$ is the periodic profile of the Wigner crystal electron density:
\begin{align}
n(\mathbf{r}) = \sum_{k,l} n_0(\mathbf{r}-k\mathbf{a}_1-l\mathbf{a}_2),
\end{align}
where $\mathbf{a}_{1,2}$ are translation vectors of the Wigner lattice. We assume the gaussian profile of the electron density:
\begin{align}
n_0(\mathbf{r}) = \frac{1}{\pi u^2} e^{-r^2/u^2},
\end{align}
where $u$ is the uncertainty of the electron position in the Wigner unit cell which is temperature- and concentration dependent. For the case of $u\ll a$ we can approximate:
\begin{align}
V_Q \approx \delta E_X(n) e^{-4\pi^2u^2/(3a^2)}.
\end{align}
The spread of the electron wavepacket, $u$ can be estimated by~\cite{monarkha2012two}:
\begin{align}
u^2/a^2\approx \frac{0.463}{\sqrt{r_s}}+0.75 r_s \frac{T}{e^2/a_B}\ln \left(\frac{L}{a_B}\right), \label{eq:wavepacket}
\end{align}
where $r_s=(a_B\sqrt{\pi}n)^{-1}$, $a_{{B}}$ is the Bohr radius, and $L$ is the lateral size of the sample. The first term corresponds to the quantum fluctuations present at zero temperature, and the second one to the thermal fluctuations of the electron density.
\subsection*{The role of disorder.}
If we assume that the phase diagram of the Wigner crystal follows the universal result of Platzman and Fukuyama~\cite{PhysRevB.10.3150}, we can use the experimental data to  extract the value of the zero temperature critical density $n_c$ corresponding to the transition from WC to Fermi liquid phase. We obtain the value of $n_c\approx 2.5\times 10^{11}$~cm${-2}$ which is more than 2 times larger than the value predicted for the WSe${_2}$~\cite{PhysRevB.95.115438}. We attribute this discrepancy to the role of the short-range disorder. Indeed, it was predicted~\cite{joy2025disorderinducedliquidsolidphasecoexistence} that the short range impurities with concentration $n_i$ and potential $\sum_{\mathbf{r'}}V_0\delta(\mathbf{r-r'})$ increase the maximum density of the Wigner phase $n_c$ with respect to to the clean limit $n_{c,0}$ as:
\begin{align}
n_c = n_{c,0}\left(1 + \frac{|\epsilon_{\mathrm{imp}}|}{|\Delta \mu(n_c)|}\right),
\end{align}
where $\epsilon_{\mathrm{imp}}= n_i V_0^2 m/\hbar^2$ is the characteristic impurity energy, and $\Delta \mu$ is the difference of energy per electron for Fermi liquid and WC phase. Since, from the experimental data $n_c\approx n_{c,0}$ we can use the results of numerical calculations for $\Delta\mu$~\cite{drummond2009phase}, to estimate $\epsilon_{\mathrm{imp}}\approx 2$~meV. Also, following~\cite{joy2025disorderinducedliquidsolidphasecoexistence} we conclude that the disorder smears the phase boundary between WC and Fermi liquid and the width of the phase coexistence region $\Delta n/n_c\sim 1$.

At the same time, the impurities lead to the disorder of the position of the nodes of the Wigner lattice which can be estimated as~\cite{joy2025disorderinducedliquidsolidphasecoexistence}:
\begin{align}
(\delta r)^2/a^2 \approx \frac{\epsilon_{\mathrm{imp}}}{4 e^2/a}.
\end{align}
The oscillator strength of the Umklapp diffraction peak will be additionally suppressed by the factor $\exp(-(4/3)\pi^2 (\delta r)^2/a^2)$. For low temperatures, this factor would substantially decrease the signal since $\delta r$ is comparable with the width of the electron wavepacket induced by the quantum fluctuations. At the same time at higher temperatures, the contribution of thermal fluctuations is dominant and the effect of disorder is less pronounced.

\section{Possible alternative origins of the additional peak.}
In this section we analyze the alternative origins of the additional peak and explain why they seem far less plausible than the Wigner phase induced Umklapp scattering.

First we can rule out the Rydberg states of the $A$ exciton as well as $B$ exciton since they are far blue shifted with respect to the observed ones~\cite{golyshkov2024excited}. Specifically $2s$ $A$ exciton states are observed at the energies larger than $1.82$ eV, and $B$ exciton at energies larger than $2$ eV. The Rydberg trion states in monolayer WSe$_2$ have also been observed~\cite{liu2021exciton} but their energies are typically above $1.8$ eV. 

Furthermore it is known~\cite{kim2024electrostatic} that encapsulating hBN may in principle induce moire-like electrostatic potential for the exciton, which could lead to the Umklapp scattering. At the same time, the reported periods of the induced potential are hundreds of nms, i.e. much larger than observed in our experiments, and furthermore these peaks should have no pronounced dependence on the applied gate voltage.

Finally, one could suggest that the observed peaks are just the consequence of the disorder. At the same time, it is not clear what should be the mechanism of the gate voltate dependence of peak energy in this case. Moreover, we have observed the peaks of very similar energy at two distinct regions in the sample which makes the hypothesis of purely disorder origin of the peaks highly implausible.


\end{document}